\documentclass[%
 reprint,
superscriptaddress,
 amsmath,amssymb,
 aps,
prd
]{revtex4-2}

\usepackage{graphicx}
\usepackage{dcolumn}
\usepackage{bm}
\usepackage{hyperref}
\hypersetup{colorlinks=true, allcolors=blue,}
\usepackage[mathlines]{lineno}
\usepackage{subfiles}
\usepackage{textgreek}
\usepackage{orcidlink}
\usepackage[separate-uncertainty=true]{siunitx}
\usepackage{tabularx}

\newcommand{\thisbegindate}{October 21, 2016}
\newcommand{\thisenddate}{November 21, 2019}
\newcommand{\totaluptime}{2.82~yr}
\newcommand{\totalexposure}{173~kg$\cdot$yr}
\newcommand{\totalmass}{61.3~kg}
\newcommand\isotope[2]{\textsuperscript{#2}#1}
\newcommand{\cosine}{\mbox{COSINE-100}}

\begin{document}

\title{Three-year annual modulation search with \cosine}

\author{G.~Adhikari}
\affiliation{Department of Physics, University of California San Diego, La Jolla, CA 92093, USA}
\author{E.~Barbosa~de~Souza}
\affiliation{Department of Physics and Wright Laboratory, Yale University, New Haven, CT 06520, USA}
\author{N.~Carlin}
\affiliation{Physics Institute, University of S\~{a}o Paulo, 05508-090, S\~{a}o Paulo, Brazil}
\author{J.~J.~Choi}
\affiliation{Department of Physics and Astronomy, Seoul National University, Seoul 08826, Republic of Korea} 
\author{S.~Choi}
\affiliation{Department of Physics and Astronomy, Seoul National University, Seoul 08826, Republic of Korea} 
\author{A.~C.~Ezeribe}
\affiliation{Department of Physics and Astronomy, University of Sheffield, Sheffield S3 7RH, United Kingdom}
\author{L.~E.~Fran{\c c}a}
\affiliation{Physics Institute, University of S\~{a}o Paulo, 05508-090, S\~{a}o Paulo, Brazil}
\author{C.~Ha}
\affiliation{Department of Physics, Chung-Ang University, Seoul 06973, Republic of Korea}
\author{I.~S.~Hahn}
\affiliation{Department of Science Education, Ewha Womans University, Seoul 03760, Republic of Korea} 
\affiliation{Center for Exotic Nuclear Studies, Institute for Basic Science (IBS), Daejeon 34126, Republic of Korea}
\affiliation{IBS School, University of Science and Technology (UST), Daejeon 34113, Republic of Korea}
\author{S.~J.~Hollick}
\affiliation{Department of Physics and Wright Laboratory, Yale University, New Haven, CT 06520, USA}
\author{E.~J.~Jeon}
\affiliation{Center for Underground Physics, Institute for Basic Science (IBS), Daejeon 34126, Republic of Korea}
\author{J.~H.~Jo}
\affiliation{Department of Physics and Wright Laboratory, Yale University, New Haven, CT 06520, USA}
\author{H.~W.~Joo}
\affiliation{Department of Physics and Astronomy, Seoul National University, Seoul 08826, Republic of Korea} 
\author{W.~G.~Kang}
\affiliation{Center for Underground Physics, Institute for Basic Science (IBS), Daejeon 34126, Republic of Korea}
\author{M.~Kauer}
\affiliation{Department of Physics and Wisconsin IceCube Particle Astrophysics Center, University of Wisconsin-Madison, Madison, WI 53706, USA}
\author{H.~Kim}
\affiliation{Center for Underground Physics, Institute for Basic Science (IBS), Daejeon 34126, Republic of Korea}
\author{H.~J.~Kim}
\affiliation{Department of Physics, Kyungpook National University, Daegu 41566, Republic of Korea}
\author{J.~Kim}
\affiliation{Department of Physics, Chung-Ang University, Seoul 06973, Republic of Korea}
\author{K.~W.~Kim}
\affiliation{Center for Underground Physics, Institute for Basic Science (IBS), Daejeon 34126, Republic of Korea}
\author{S.~H.~Kim}
\affiliation{Center for Underground Physics, Institute for Basic Science (IBS), Daejeon 34126, Republic of Korea}
\author{S.~K.~Kim}
\affiliation{Department of Physics and Astronomy, Seoul National University, Seoul 08826, Republic of Korea}
\author{W.~K.~Kim}
\affiliation{IBS School, University of Science and Technology (UST), Daejeon 34113, Republic of Korea}
\affiliation{Center for Underground Physics, Institute for Basic Science (IBS), Daejeon 34126, Republic of Korea}
\author{Y.~D.~Kim}
\affiliation{Center for Underground Physics, Institute for Basic Science (IBS), Daejeon 34126, Republic of Korea}
\affiliation{Department of Physics, Sejong University, Seoul 05006, Republic of Korea}
\affiliation{IBS School, University of Science and Technology (UST), Daejeon 34113, Republic of Korea}
\author{Y.~H.~Kim}
\affiliation{Center for Underground Physics, Institute for Basic Science (IBS), Daejeon 34126, Republic of Korea}
\affiliation{Korea Research Institute of Standards and Science, Daejeon 34113, Republic of Korea}
\affiliation{IBS School, University of Science and Technology (UST), Daejeon 34113, Republic of Korea}
\author{Y.~J.~Ko}
\affiliation{Center for Underground Physics, Institute for Basic Science (IBS), Daejeon 34126, Republic of Korea}
\author{H.~J.~Kwon}
\affiliation{IBS School, University of Science and Technology (UST), Daejeon 34113, Republic of Korea}
\affiliation{Center for Underground Physics, Institute for Basic Science (IBS), Daejeon 34126, Republic of Korea}
\author{D.~H.~Lee}
\affiliation{Department of Physics, Kyungpook National University, Daegu 41566, Republic of Korea}
\author{E.~K.~Lee}
\affiliation{Center for Underground Physics, Institute for Basic Science (IBS), Daejeon 34126, Republic of Korea}
\author{H.~Lee}
\affiliation{IBS School, University of Science and Technology (UST), Daejeon 34113, Republic of Korea}
\affiliation{Center for Underground Physics, Institute for Basic Science (IBS), Daejeon 34126, Republic of Korea}
\author{H.~S.~Lee}
\affiliation{Center for Underground Physics, Institute for Basic Science (IBS), Daejeon 34126, Republic of Korea}
\affiliation{IBS School, University of Science and Technology (UST), Daejeon 34113, Republic of Korea}
\author{H.~Y.~Lee}
\affiliation{Center for Underground Physics, Institute for Basic Science (IBS), Daejeon 34126, Republic of Korea}
\author{I.~S.~Lee}
\affiliation{Center for Underground Physics, Institute for Basic Science (IBS), Daejeon 34126, Republic of Korea}
\author{J.~Lee}
\affiliation{Center for Underground Physics, Institute for Basic Science (IBS), Daejeon 34126, Republic of Korea}
\author{J.~Y.~Lee}
\affiliation{Department of Physics, Kyungpook National University, Daegu 41566, Republic of Korea}
\author{M.~H.~Lee}
\affiliation{Center for Underground Physics, Institute for Basic Science (IBS), Daejeon 34126, Republic of Korea}
\affiliation{IBS School, University of Science and Technology (UST), Daejeon 34113, Republic of Korea}
\author{S.~H.~Lee}
\affiliation{IBS School, University of Science and Technology (UST), Daejeon 34113, Republic of Korea}
\affiliation{Center for Underground Physics, Institute for Basic Science (IBS), Daejeon 34126, Republic of Korea}
\author{S.~M.~Lee}
\affiliation{Department of Physics and Astronomy, Seoul National University, Seoul 08826, Republic of Korea} 
\author{D.~S.~Leonard}
\affiliation{Center for Underground Physics, Institute for Basic Science (IBS), Daejeon 34126, Republic of Korea}
\author{B.~B.~Manzato}
\affiliation{Physics Institute, University of S\~{a}o Paulo, 05508-090, S\~{a}o Paulo, Brazil}
\author{R.~H.~Maruyama}
\affiliation{Department of Physics and Wright Laboratory, Yale University, New Haven, CT 06520, USA}
\author{R.~J.~Neal}
\affiliation{Department of Physics and Astronomy, University of Sheffield, Sheffield S3 7RH, United Kingdom}
\author{B.~J.~Park}
\affiliation{IBS School, University of Science and Technology (UST), Daejeon 34113, Republic of Korea}
\affiliation{Center for Underground Physics, Institute for Basic Science (IBS), Daejeon 34126, Republic of Korea}
\author{H.~K.~Park}
\affiliation{Department of Accelerator Science, Korea University, Sejong 30019, Republic of Korea}
\author{H.~S.~Park}
\affiliation{Korea Research Institute of Standards and Science, Daejeon 34113, Republic of Korea}
\author{K.~S.~Park}
\affiliation{Center for Underground Physics, Institute for Basic Science (IBS), Daejeon 34126, Republic of Korea}
\author{S.~D.~Park}
\affiliation{Department of Physics, Kyungpook National University, Daegu 41566, Republic of Korea}
\author{R.~L.~C.~Pitta}
\affiliation{Physics Institute, University of S\~{a}o Paulo, 05508-090, S\~{a}o Paulo, Brazil}
\author{H.~Prihtiadi}
\affiliation{Center for Underground Physics, Institute for Basic Science (IBS), Daejeon 34126, Republic of Korea}
\author{S.~J.~Ra}
\affiliation{Center for Underground Physics, Institute for Basic Science (IBS), Daejeon 34126, Republic of Korea}
\author{C.~Rott}
\affiliation{Department of Physics, Sungkyunkwan University, Suwon 16419, Republic of Korea}
\affiliation{Department of Physics and Astronomy, University of Utah, Salt Lake City, UT 84112, USA}
\author{K.~A.~Shin}
\affiliation{Center for Underground Physics, Institute for Basic Science (IBS), Daejeon 34126, Republic of Korea}
\author{A.~Scarff}
\affiliation{Department of Physics and Astronomy, University of Sheffield, Sheffield S3 7RH, United Kingdom}
\author{N.~J.~C.~Spooner}
\affiliation{Department of Physics and Astronomy, University of Sheffield, Sheffield S3 7RH, United Kingdom}
\author{W.~G.~Thompson\,\orcidlink{0000-0003-2988-7998}}
\email[Corresponding author: ]{william.thompson@yale.edu}
\affiliation{Department of Physics and Wright Laboratory, Yale University, New Haven, CT 06520, USA}
\author{L.~Yang}
\affiliation{Department of Physics, University of California San Diego, La Jolla, CA 92093, USA}
\author{G.~H.~Yu}
\affiliation{Department of Physics, Sungkyunkwan University, Suwon 16419, Republic of Korea}
\collaboration{COSINE-100 Collaboration}
\noaffiliation

\date{\today}

\begin{abstract}
\cosine\ is a direct detection dark matter experiment that aims to test DAMA/LIBRA's claim of dark matter discovery by searching for a dark matter-induced annual modulation signal with NaI(Tl) detectors. We present new constraints on the annual modulation signal from a dataset with a \totaluptime\ livetime utilizing an active mass of \totalmass\, for a total exposure of \totalexposure. This new result features an improved event selection that allows for both lowering the energy threshold to 1~keV and a more precise time-dependent background model. In the 1--6~keV and 2--6~keV energy intervals, we observe best-fit values for the modulation amplitude of 0.0067$\pm$0.0042 and 0.0051$\pm$0.0047~counts/(day$\cdot$kg$\cdot$keV), respectively, with a phase fixed at 152.5~days. 
\end{abstract}

\maketitle
\section{Introduction}
\label{intro}

Cosmological observations indicate that more than a quarter of our Universe's mass-energy exists in the form of a massive, non-luminous component known as dark matter~\cite{Planck2020}. The apparent abundance of dark matter has given rise to several experiments over the past three decades that aim to observe dark matter directly~\cite{pandax_2021,XENON1T,LUX-complete,PICO-60,XMASS2018,deap3600_2019,darkside50_2018,supercdms_2018,KIMS}. Despite this large-scale effort, none of these experiments have seen any signal indicating the existence of dark matter, except for the DAMA experiments, DAMA/NaI~\cite{Bernabei:2005hj} and DAMA/LIBRA~\cite{DAMAFinal,DAMAPhase2}.

DAMA's claim of dark matter discovery comes in the form of an annual modulation in the event rate of thallium-doped sodium iodide (NaI(Tl)) detectors, which has persisted for more than two decades. This observed modulation possesses the characteristics expected of a dark matter-induced annual modulation signal~\cite{modulation,Colloquium}, including a maximum rate near June 2\textsuperscript{nd} and a period of one year. Additionally, this modulation signal is observed at an extremely high significance of 12.9$\sigma$ in the 2--6~keV energy range and 9.5$\sigma$ in the 1--6~keV energy range compared with the no-modulation hypothesis~\cite{DAMAPhase2}. Despite this high significance, the DAMA result is in severe tension with the null results obtained by other direct-detection dark matter experiments within most commonly considered models of dark matter~\cite{Schumann_2019}.

The \cosine\ collaboration aims to resolve this tension by performing a model-independent test of the DAMA collaboration's claim of dark matter discovery~\cite{COSINE_detector}. This model independence is achieved by using the same target material as the DAMA experiments, NaI(Tl). Previously, we have excluded spin-independent WIMP-nucleus interactions as the origin of DAMA's modulation signal within the context of the standard halo model, becoming the first NaI(Tl)-based experiment to test the DAMA result~\cite{COSINE_WIMP,cosine_wimp_set2}. We have also published results on a model-independent search for the DAMA-observed modulation with the initial 1.7~yr of data from \cosine~\cite{COSINE_modulation}. This first modulation search was statistically limited and thus compatible with both the no annual modulation hypothesis and the DAMA experiments' best-fit modulation amplitude in the 2--6~keV region. In addition, the \mbox{ANAIS-112} collaboration, which also aims to test DAMA's discovery claim with an array of NaI(Tl) detectors, has released results based on one and a half~\cite{ANAIS2019} and three years of detector operation~\cite{ANAIS_threeyears}, the latter of which is incompatible with the DAMA/LIBRA result at 3.3$\sigma$. 

In this paper, we present the results of a search for a dark matter-induced annual modulation signal in \cosine\ from \thisbegindate\ to \thisenddate, with a total livetime of \totaluptime. In addition to the increased exposure, this analysis features a number of improvements to our previous modulation search, including a decrease in energy threshold to 1~keV~\cite{cosine_low_threshold}, a time-dependent background model based on our modeling reported in~\cite{COSINE_background2}, and the implementation of a Bayesian analysis approach.

We first provide, in Section~\ref{det}, an overview of the \cosine\ detector and its performance over the three-year period of data-taking investigated in this analysis. Section~\ref{sec:event_selection} details the updated event selection used to lower our energy threshold to 1~keV. In Section~\ref{sec:bkgd}, we describe the background model used in this analysis. In Section~\ref{sec:modulation}, we present our modulation search procedure and results from this analysis, with concluding remarks given in Section~\ref{conclusions}.

\section{Detector Overview and Stability}
\label{det}

\subsection{Detector Design}

A full description of the \cosine\ detector has been previously published in Refs.~\cite{COSINE_detector,thesis_thompson}. In this section, we provide a brief overview of aspects of the detector configuration and performance that are relevant to the annual modulation search.

The \cosine\ detector consists of eight low-background NaI(Tl) detectors, with a total mass of 106~kg, located at a depth of approximately 1800~meters water equivalent in the Yangyang underground lab in South Korea. These detectors are submerged in a 2200~L liquid scintillator (LS) detector that serves as a background veto system~\cite{COSINE_LS}. This veto reduces backgrounds from radiation external to the NaI(Tl) detectors, in addition to enabling the tagging and removal of internal low-energy backgrounds that are accompanied by a high-energy emission, particularly decays from \isotope{K}{40}. The LS is continually flushed with nitrogen gas in order to remove radon from the detector volume and prevent discoloring of the scintillator from contact with oxygen. External backgrounds are further reduced by a shielding structure composed of a layer of 3-cm thick copper surrounded by 20~cm of lead. Thirty-seven 3-cm thick plastic scintillator panels placed around the shielding structure provide 4$\pi$ cosmic ray muon tagging~\cite{COSINE_muon_det,cosine_muon_measurement}.

Each NaI(Tl) crystal detector is optically coupled to two photomultiplier tubes (PMTs) that detect scintillation photons from energy depositions in the detector. These eight detectors are referred to as Crystal 1 (C1) through Crystal 8 (C8). C1, C5, and C8 are excluded from this analysis due to high background levels and low light yields in the case of C5 and C8, and a high noise rate in the case of C1. This  results in a total effective mass of \totalmass.

\subsection{Detector Stability}
\label{stability}

The \cosine\ detector room is instrumented with sensors that monitor the environmental conditions of the room and the detector. This includes sensors that monitor the room temperature, humidity, and radon level, and the temperature of the LS volume, which serves as a proxy for the NaI(Tl) detectors' temperatures. Continuous monitoring of these environmental conditions allows us to assess the stability of the environment and detector in real-time. The monitoring system also allows us to perform offline investigations into possible correlations between environmental conditions and the detector event rate. A detailed description of the monitoring system can be found in Ref.~\cite{cosine_env_monitor}.

We calibrate the NaI(Tl) detectors using both external $\gamma$-ray sources and $\beta$- and $\gamma$-ray emissions from radioactive contaminants internal to the NaI(Tl) detectors. At energies below 70~keV, the nature of the non-linear light response of NaI(Tl) is studied using low-energy internal decays. We account for this non-linearity in our energy calibration via an empirical model based on these low-energy decays. Full details of the energy calibration method can be found in Ref.~\cite{COSINE_background2}.

In addition to accounting for non-linearities in the light yield at low energies, it is necessary to monitor and correct for gain shifts in the NaI(Tl)-coupled PMTs to ensure a stable detector energy scale. Stability of the energy scale is of particular importance for annual modulation searches, as instabilities can manifest as changes in the event rate, possibly inducing an artificial annual modulation signal. To guard against this effect, we monitor and correct for gain shifts in each of the NaI(Tl)-coupled PMTs by tracking changes in the reconstructed energy of the $\sim$50~keV decay from \isotope{Pb}{210}. This decay line originates from the combination of an electron emitted in the $\beta$-decay of \isotope{Pb}{210} to \isotope{Bi}{210} ($Q = 17$~keV) and a prompt 46.5~keV $\gamma$-ray from the corresponding relaxation of the excited \isotope{Bi}{210} nucleus. Since this decay originates from radioisotopes internal to the NaI(Tl) detector, it enables continuous monitoring of the PMT gains over the course of the data run. In practice, we measure the reconstructed energy of this decay peak over contiguous 20-hour periods over the full course of the run. The average size of the gain shift across the NaI(Tl)-coupled PMTs is $<10\%$ over the three-year-long dataset and is less than 15\% in all PMTs. To stabilize each NaI(Tl) detector, we apply a time-dependent correction factor to the energy scale of each PMT based on the observed evolution of the reconstructed energy of the \isotope{Pb}{210} peak over time.

\begin{figure}[t]
	\centering
	\includegraphics[width=\columnwidth]{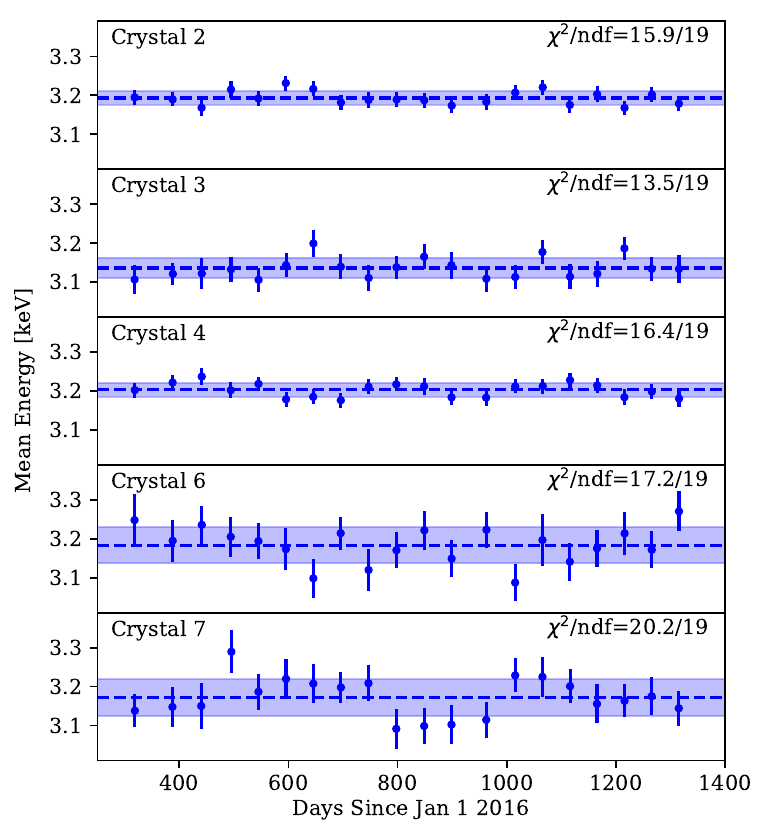}
	\caption{Energy scale stability over time in the 1--6 keV region of interest for the five detectors used in this analysis. Blue data points represent the mean energy of the fitted 3.2~keV x-ray peaks from \isotope{K}{40}. The dashed lines are the mean values of the peak positions in each NaI(Tl) detector, and the shaded regions denote the standard deviations of the peak positions. Crystals 6 and 7 have the lowest \isotope{K}{40} contamination, resulting in larger uncertainties in the peak positions when compared with the other detectors.}
    \label{fig:stability}
\end{figure}

After the application of this correction, we assess the stability of each gain-corrected NaI(Tl) detector in the 1--6~keV region of interest (ROI) by tracking the measured energy of the 3.2~keV x-ray from the decay of \isotope{K}{40} within each sodium iodide detector over time. As this 3.2~keV x-ray is emitted in coincidence with a higher-energy 1461~keV $\gamma$-ray, we are able to select a high-purity sample of 3.2~keV energy depositions in each NaI(Tl) detector by requiring a coincident energy deposition between 900 and 1550~keV within our LS veto. With the gain-shift correction applied, we find that the energy scale over time is stable within the ROI in all five NaI(Tl) detectors used in this analysis, as illustrated in Fig.~\ref{fig:stability}. Specifically, we find no evidence of a change in any of the detectors' energy scales over time, with a maximum reduced chi-square statistic of $\chi^2/$DOF$=20.2/19$ across all five detectors. 

\section{Event Selection}
\label{sec:event_selection}

In this study, we employ a newly developed event selection procedure that enables lowering the analysis energy threshold to 1~keV, an improvement compared with the 2~keV threshold in our previous modulation search~\cite{COSINE_modulation}. A full description of the updated event selection procedure can be found in Ref.~\cite{cosine_low_threshold}. Here, we provide a summary of this procedure.

The first step in the event selection procedure is to identify and remove muon-induced events, defined as any event that occurs within 30~ms of an energy deposition in the muon veto system. We also remove events with characteristics that are consistent with noise signals that originate from electronic pickup. Next, we classify the remaining events as ``single-hit" or ``multi-hit." Specifically, a single-hit event is any event in which there is an energy deposition in only a single NaI(Tl) detector along with no measurable energy deposition in the LS veto; a multi-hit event is any event with an energy deposition in multiple NaI(Tl) detectors, or an event with an energy deposition in at least one NaI(Tl) detector and the LS veto. Dark matter, due to its extremely small probability of interacting with ordinary matter, is expected to exclusively induce single-hit events. While multi-hit events are not directly used in the dark matter search, they provide a convenient sideband sample for the evaluation of cut efficiencies and the assessment of other properties of our modulation search procedure.

The primary focus of our event-selection procedure is to remove PMT-induced noise events that mimic the shape of NaI(Tl) scintillation signals. At lower energies, it becomes increasingly difficult to distinguish between signal and PMT-induced noise events due to the fewer number of photons emitted by a NaI(Tl) detector. Thus, decreasing our energy threshold required the development of a more powerful noise discrimination algorithm than that of our previous approach~\cite{COSINE_WIMP}. This updated event selection procedure utilizes a new metric, named the ``likelihood score," that compares individual PMT waveforms to template signal and noise waveforms. The signal template is obtained from a high-purity calibration dataset of scintillation events. The events in this high-purity dataset originate from Compton-scattered $\gamma$-rays from a \isotope{Co}{60} calibration source and were isolated from non-scintillation noise events by selecting only multi-hit events. The noise template is created using events identified as noise via an independent pulse-shape identification procedure. The difference between the signal and noise likelihood values is defined as the likelihood score for each individual event.

We utilize a boosted decision tree (BDT) to combine the likelihood score with the event energy and other event discrimination features that were used in our previous modulation search~\cite{COSINE_modulation}. These other features are used to further quantify PMT waveforms shapes, as well as the asymmetry of light collection between the two PMTs coupled to a single NaI(Tl) detector. Though the precise weighting of each feature by the BDT varies between the five detectors, their general order of importance is the same. The likelihood parameter is assigned the largest weight by the BDT, followed by the event energy. The asymmetry feature is the third highest weighted, with high asymmetry values being characteristic of noise events. The asymmetry feature is followed in importance by parameters characterizing the prompt and delayed charge measured by each of the two PMTs attached to a given NaI(Tl) detector. The other waveform shape parameters used in previous analyses by \cosine\ make up the remaining, lower-ranked parameters~\cite{COSINE_WIMP,COSINE_modulation}. The high ranking of the likelihood parameter illustrates its importance in the updated event selection procedure used in this analysis.

To compute the efficiency of the event selection as a function of energy, we again utilize a high-purity dataset obtained from a \isotope{Co}{60} calibration run. The selection efficiency is defined as the ratio between the number of Compton-scatter events after to the number of Compton-scatter events before the selection and is computed independently for each of the five detectors used in this analysis. We find that this updated event selection achieves a signal selection efficiency above 80\% in the 1--1.5~keV energy bin and that approaches unity above 2~keV~\cite{cosine_wimp_set2}. For this analysis, the selection efficiency uncertainty of each NaI(Tl) detector was obtained from the statistical uncertainties of the \isotope{Co}{60} dataset. Because of the known difference in waveform shape between electronic and nuclear recoil events in NaI(Tl) detectors~\cite{gerbier_1999}, we also measure the efficiency of the signal selection on low-energy nuclear recoil events obtained by irradiating a small NaI(Tl) detector with 2.42~MeV neutrons from a D-D neutron generator~\cite{joo_2019}. This cross-check is of particular importance given that dark matter is expected to interact with NaI(Tl) detectors by inducing nuclear recoils. The selection efficiencies measured with the nuclear recoil dataset are consistent with those measured using the \isotope{Co}{60} dataset~\cite{cosine_wimp_set2}, verifying that the event selection does not preferentially remove events generated by dark matter interactions.

Lastly, two-hour periods of data are occasionally removed from consideration in our analysis due to environmental or detector instability. Any large variation in the environmental or detector parameters recorded results in the removal of the corresponding two-hour-long sub-run. One example of such an instability is a discrete spike in the detector trigger rate due to the passage of a high-energy cosmic ray muon through the detector; in this case the corresponding sub-run would be removed to mitigate any possible impact of long-lived muon-induced phosphorescence. In addition, there were a few occasions on which a NaI(Tl) detector experienced a large increase in the number of PMT-induced noise events, resulting in significantly elevated event rates over a sustained period of time~\cite{COSINE_modulation,cosine_wimp_set2}. Because these periods of high event rates are not correlated between NaI(Tl) detectors, exclusion of a particular sub-run from the analysis occurs on a detector-by-detector basis. While we expect the event selection procedure to remove the large majority of these events, we nonetheless have fully removed these periods of instability from consideration in this analysis to preclude the possibility of observing a noise-induced modulation signal. The main periods affected by these elevated noise rates are from October 21 to December 19, 2016 and July 27 to August 8, 2019 for C2, and January 12 to March 31, 2017 for C7. After accounting for the removal of these sub-runs and detector calibration periods, we achieve a 91.5\% livetime with the \cosine\ detector from the operating period spanning from \thisbegindate\ to \thisenddate, resulting in a total data exposure of \totalexposure. Additionally, 5\% of the time over this period is composed of deadtime, which is defined as both periods during which the \cosine\ detector was not recording data and periods excluded due to detector instability. Lastly, detector calibrations comprise 3.5\% of the detector operating period considered in this analysis.

\section{Background Model}
\label{sec:bkgd}

As the search for an annual modulation signal is primarily focused on the \cosine\ detector's time-dependent event rate, it is vital to understand any time-dependent background components that can affect the signal rate in our 1--6~keV ROI. Mismodeling these time-dependent background components can bias estimates of the amplitude of the sought annual modulation signal, as has been discussed in Buttazzo \textit{et al}.~\cite{Buttazzo2020} and Messina \textit{et al}~\cite{Messina2020}.

\begin{figure}[!ht]
	\centering
	\includegraphics[width=\columnwidth]{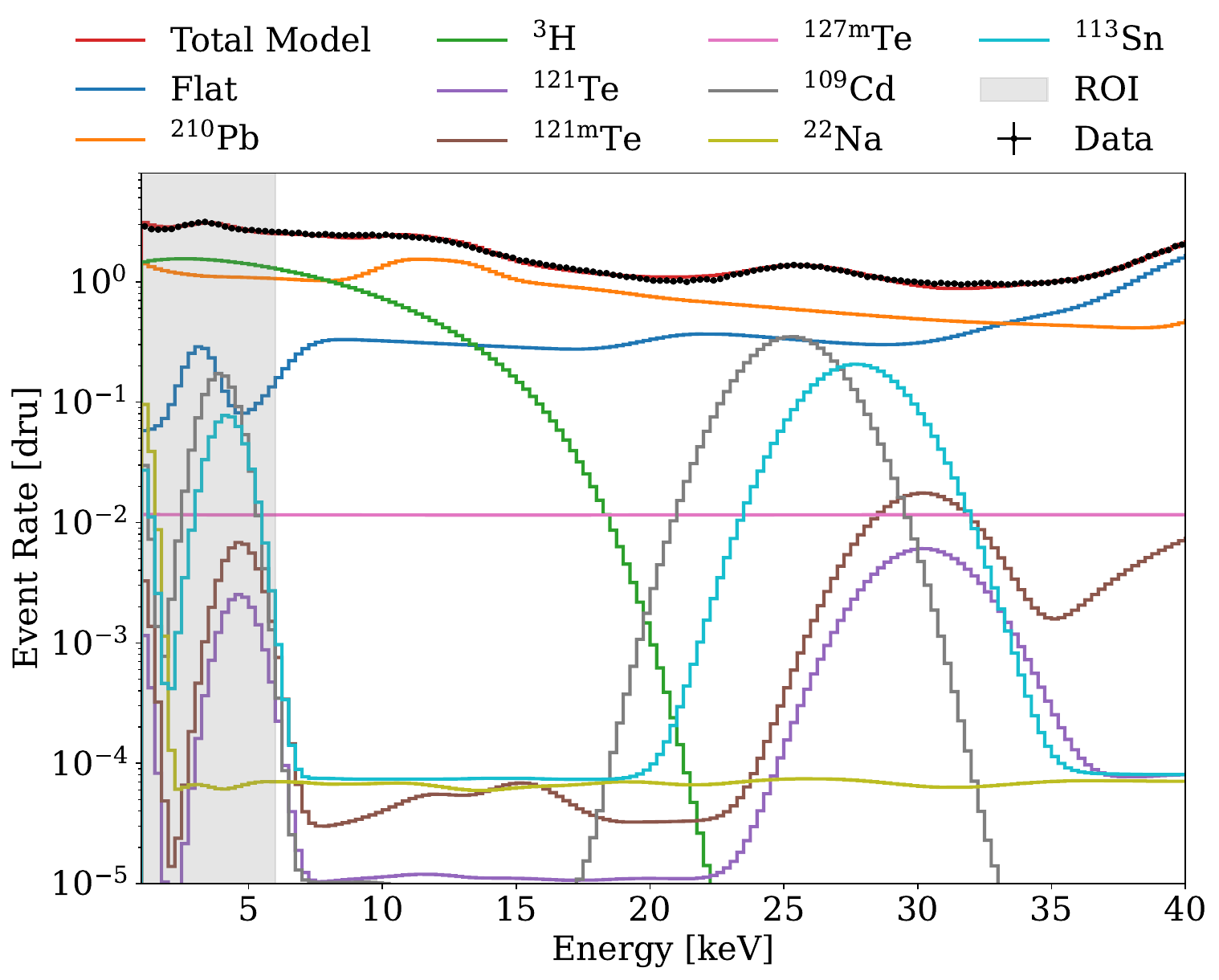}
	\caption{Best-fit background model compared with observed data (black histogram) averaged across all five NaI(Tl) detectors used in this analysis from Ref.~\cite{COSINE_background2}. The red histogram is the total energy spectrum from our background model. Isotopes with activities modeled as constant in time in our analysis are grouped into the ``flat" component (blue histogram). The energy spectra of isotopes whose time dependence is considered within our annual modulation search model are shown individually. The gray shaded region denotes the region of interest for the annual modulation search.}
	\label{fig:bkgd_model}
\end{figure}

A primary focus of the analysis reported here is the development of a background model that fully accounts for each of these time-dependent components, improving upon the background model used in our previous modulation search~\cite{COSINE_modulation}.  Primarily, these short-lived components are cosmogenic radioisotopes created by the interaction of sodium and iodine nuclei with high-energy cosmic rays before the NaI(Tl) detectors were brought underground.
We identified particular cosmogenic isotopes expected to contribute to the time-dependent background of the experiment based on prior studies of cosmogenically activated NaI(Tl) detectors~\cite{thesis_pettus,anais_cosmo_2015,anais_cosmo_2018,anais_tritium_2018}. We have previously reported an independent measurement of these cosmogenic isotopes' activities~\cite{COSINE_cosmogenics} and incorporated this information into the full, time-integrated background model of \cosine, as detailed in Ref.~\cite{COSINE_background2}. Additionally, we have improved the \cosine\ background model by refining the spatial distribution of \isotope{Pb}{210} contamination in our NaI(Tl) detectors, also detailed in Ref.~\cite{COSINE_background2}. As \isotope{Pb}{210} is the primary background component in the majority of the NaI(Tl) detectors, it is an isotope of particular importance to our annual modulation search.

\cosine's time-integrated background model was developed by first simulating the predicted energy spectrum from each radioisotope expected to contaminate the detector using Geant4~\cite{geant}. The activity of each isotope was then determined by jointly fitting the amplitude of the simulated energy spectra to the energy spectrum measured by a given NaI(Tl) detector. These fits were performed separately for each detector, resulting in independent background models for each detector. Full details of the time-integrated background model are described in Ref.~\cite{COSINE_background2}. 

Fig.~\ref{fig:bkgd_model} displays the total time-integrated background model averaged across the five detectors used in this analysis in ``daily rate units," or dru, a common shorthand for counts/(day$\cdot$kg$\cdot$keV). The energy spectra of specific isotopes of interest, defined below, are also shown. Worthy of particular note is \isotope{H}{3}, which is the most prominent background component in the ROI in two of the NaI(Tl) detectors used in this analysis; \isotope{Pb}{210} is the most prominent background in the other three detectors. L-shell emissions from cosmogenic isotopes also contribute significantly to the detector background as seen in the figure. The remaining isotopes, whose activities do not vary appreciably on a three-year timescale, are grouped together in a constant-in-time, or ``flat," component. The flat component consists primarily of decays of \isotope{K}{40} and isotopes belonging to the \isotope{U}{238} and \isotope{Th}{232} decay chains.

\begin{table}
\sisetup{retain-zero-exponent = true}
\caption{Components considered in the time-dependent background model and their average activities over the course of the dataset used in the three-year modulation search.}
\label{tab:avg_bkgds}
\begin{ruledtabular}
\begin{tabular}{lr}
Component & \multicolumn{1}{l}{Average Activity (dru)} \\
\colrule
Total     & \SI{2.74(23)E+00}{} \\
\isotope{H}{3}        & \SI{1.41(18)E+00}{} \\
\isotope{Pb}{210}     & \SI{1.12(15)E+00}{} \\
Flat      & \SI{1.35(8)E-01}{} \\
\isotope{Cd}{109}     & \SI{4.13(39)E-02}{} \\
\isotope{Sn}{113}     & \SI{1.55(16)E-02}{} \\
\isotope{Te}{127}     & \SI{6.59(52)E-03}{} \\
\isotope{Na}{22}      & \SI{5.88(134)E-03}{} \\
\isotope{Te}{121m}    & \SI{1.50(16)E-03}{} \\
\isotope{Te}{121}     & \SI{5.07(123)E-04}{}
\end{tabular}
\end{ruledtabular}
\end{table}

The time-dependent background model is derived directly from this time-integrated background model. First, we determine the average event rate of each time-dependent background component within a given energy range by integrating its simulated spectrum over said energy range. Next, we compute the event rate at the beginning of the dataset considered in this analysis induced by each time-dependent component using its average event rate, half-life, and livetime history. From this calculation, we identify eight radioisotopes that are expected to have a measurable impact on the time-dependent event rate of \cosine. These are defined as isotopes whose decays induced $>$0.01\% of the total event rate in any of the five NaI(Tl) detectors on \thisbegindate, which corresponds to an event rate of $\gtrsim$\SI{3e-4}~dru. We also required that these isotopes have short half-lives, conservatively defined as less than one million years. The specific isotopes identified and their average activities over the course of the dataset used in the three-year modulation search are reported in Table~\ref{tab:avg_bkgds}. Seven of these isotopes are of cosmogenic origin, highlighting the importance of their inclusion in our background model. We define the total time-dependent background model of each NaI(Tl) detector as the sum of the constant background component and the eight exponentially decaying background components, with the activity of each time-dependent component on \thisbegindate\ treated as a nuisance parameter in the annual modulation search.

\section{Annual Modulation Search}
\label{sec:modulation}

\subsection{Modulation Search Procedure}
\label{sec:fit}

We define the total time-dependent event rate of the $i$\textsuperscript{th} NaI(Tl) detector as the sum of the event rate predicted by the time-dependent background model added to a dark matter-induced, annually modulating signal component
\begin{equation}\label{eq:fit}
	R^i(t|S_m,\alpha^i,\beta^i_k) = \alpha^i + \sum_{k=1}^{N_{bkgd}} \beta_k^{i} e^{-\lambda_k t} + S_m\cos{(\omega(t-t_{0}))} .
\end{equation}
In Eq.~\ref{eq:fit}, the sum over $k$ represents the sum over the $N_{bkgd}=8$ background components with decay constants $\lambda_k$. $\alpha^i$ is the rate from the constant background component and $\beta_k^i$ is the initial rate from the $k$\textsuperscript{th} background component, both in the $i$\textsuperscript{th} detector. The modulation is described by its amplitude $S_m$, phase $t_0$, and period $T=365.25$ days, where $\omega=2\pi/T$. Our model requires the same modulation signal amongst all detectors, but allows for different background activity levels to account for the different contamination levels across different detectors.

In order to efficiently treat the large number of nuisance parameters in our model from these background components we utilize a Bayesian analysis framework, making use of Markov chain Monte Carlo techniques to calculate the marginalized posterior distributions of the parameters of interest. The posterior distribution is given by
\begin{equation}\label{eq:post}
    P(S_m|\mathbf{x}) = N\int d\boldsymbol{\alpha} \int d\boldsymbol{\beta}\  \mathcal{L}(\mathbf{x}|S_m,\boldsymbol{\alpha},\boldsymbol{\beta})\ \pi(S_m,\boldsymbol{\alpha},\boldsymbol{\beta}),
\end{equation}
where $N$ is a normalization constant, $\pi(S_m,\boldsymbol{\alpha},\boldsymbol{\beta})$ represents the prior distributions, and $\boldsymbol{\alpha}$ and $\boldsymbol{\beta}$ are vectors whose components represent the activities of the constant and time-dependent backgrounds in each crystal, respectively. The observed data is denoted by $\mathbf{x}$ and comprises the efficiency- and livetime-corrected event rate in each time bin of each NaI(Tl) detector.

To obtain the efficiency- and livetime-corrected event rate over time of a given detector, we first compute the number of events in each 15-day time bin after application of the event selection procedure described in Sec.~\ref{sec:event_selection}. We then evaluate the detector livetime in each time bin and normalize each bin based on its relative exposure. This process accounts for variations in exposure induced by both detector-off periods and data periods removed due to detector instability. We next scale the event rate in each time bin by the reciprocal of the measured selection efficiency to obtain the sought corrected event rate over time. The uncertainty of each time bin is defined as the square root of the number of events in each time bin before correction, scaled by the efficiency and livetime corrections. Lastly, we found that the uncertainties of the measured selection efficiencies had a negligible impact on the measured modulation amplitudes and phases; they are therefore excluded from the remainder of the analysis.

We utilize a binned likelihood built as the product of Gaussian probabilities
\begin{equation}
    \mathcal{L}(\mathbf{x}|S_m,\boldsymbol{\alpha},\boldsymbol{\beta}) = \prod_{i}^{N_{det}} \prod_{j}^{N^i_{bin}}\ 
    \exp\left[ -\frac{1}{2} \left(\frac{x_{ij}-\mu_{ij}}{\sigma_{ij}} \right)^2 \right],
\end{equation}
where $\mu_{ij}$ is the expected number of counts in the $j$\textsuperscript{th} time bin of the $i$\textsuperscript{th} detector, obtained by integrating $R^i(t|S_m,\alpha^i,\beta^i)$ over the duration of the $j$\textsuperscript{th} time bin, and $\sigma_{ij}$ is its associated uncertainty. $N_{det}=5$ is the number of detectors used in this analysis, and $N^i_{bin}$ is the number of time bins in the $i$\textsuperscript{th} detector.

We assume Gaussian priors for the initial activities of all background components, with the mean and standard deviation of each component's priors set to its computed activity and its associated uncertainty on \thisbegindate, as described in Sec.~\ref{sec:bkgd}. In this paper, we summarize the marginalized posterior distributions of our parameters of interest, the modulation amplitude and phase, by reporting their highest-density credible intervals (HDIs) for one-dimensional posteriors or highest-density credible regions (HDRs) for two-dimensional posteriors.

In our primary analysis, we fix the phase of the modulation such that the maximum occurs on June 2\textsuperscript{nd}, 152.5~days from the start of the calendar year, as predicted in the standard halo model~\cite{modulation}, and search for an annual modulation signal in the 1--6~keV and 2--6~keV ROIs with a flat prior for the modulation amplitude. We also search for an annual modulation signal by allowing both the amplitude and phase of the signal to vary in the fit, with both assigned flat priors. Lastly, we also perform an analysis of the modulation amplitude as a function of energy by dividing the data into 1-keV wide bins from 1--20 keV and fitting for the modulation amplitude in each bin with the phase of the modulation signal again fixed to 152.5~days; this analysis is also performed on the multi-hit data set. As we expect to observe no significant modulation in either the 6--20~keV single-hit or 1--6~keV multi-hit datasets, these datasets serve as sideband regions to validate our fitting procedure and background model.

\subsection{Pseudo-Experiment Validation}
\label{sec:pseudo}

To examine the validity of our fitting procedure, we generate five pseudo-experiment ensembles that are analyzed using the same fitting procedure as is applied to the measured data. For a single pseudo-experiment, the event rate over time in each NaI(Tl) detector is generated from our signal plus background model described by Eq.~\ref{eq:fit} with Poissonian fluctuations introduced in each 15-day time bin. Each ensemble consists of 1000 pseudo-experiments. The initial activities of each background component are determined from our background model. In order to consider the possible parameter space of initial activities of the background components in our model, we randomly select the initial activities of each component from their respective prior distributions. The amplitude of the injected modulation signal is varied between the five different ensembles to elucidate characteristics of our modulation search procedure across an amplitude range that extends from 0.0000 to 0.0105~dru.

\begin{figure}[t]
	\centering
	\includegraphics[width=\columnwidth]{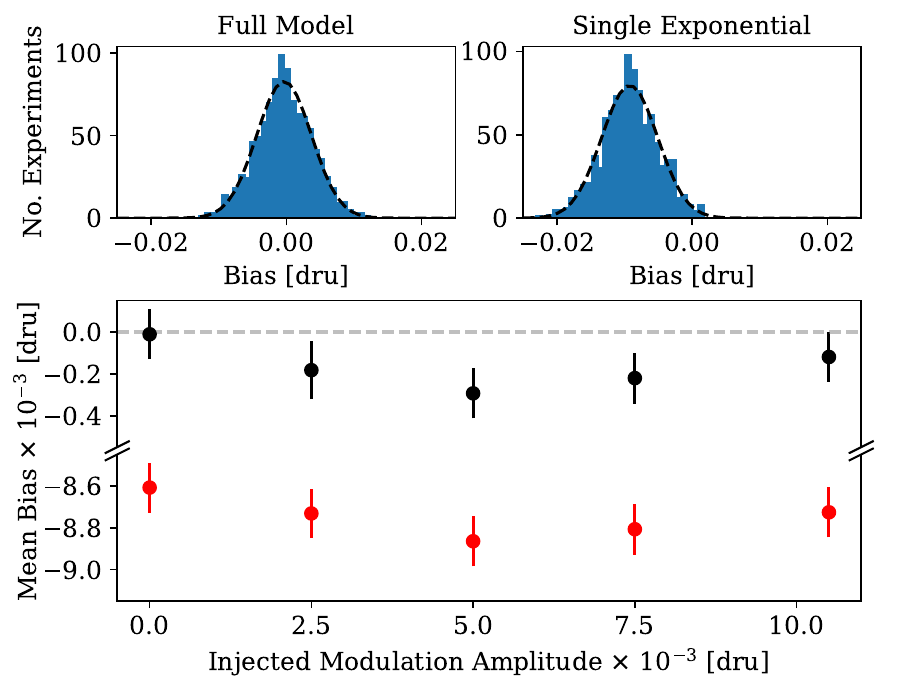}
	\caption{Summary of results from the single-hit, phase-fixed pseudo-experiment studies. The upper plots show, for the no-modulation pseudo-experiment ensemble, the bias distributions of the fitted amplitudes for the model used in this analysis and described in Eq.~\ref{eq:fit} (left) and for the single-exponential model (right). Results of Gaussian fits to these bias distributions are also shown (black dashed lines). The bottom plot shows the means of the fitted Gaussian distributions as a function of the simulated modulation amplitudes; the fits to the full model bias distributions are in black and to the single-exponential bias distributions are in red.}
    \label{fig:pseudo}
\end{figure}

We quantify the validity of our search procedure by investigating the bias distribution of the fitted modulation amplitudes of each pseudo-experiment ensemble. Each bias distribution is fitted with a Gaussian function, from which the mean bias of the ensemble is determined. At the modulation amplitudes investigated, we find a maximum bias in our search procedure of -0.0003~dru, which occurs for a 0.0050~dru input modulation amplitude. The results of this investigation are summarized in Fig.~\ref{fig:pseudo}. Additionally, we utilize these pseudo-experiment ensembles to evaluate the projected uncertainty of the measured amplitude for modulations of various sizes. This is achieved by investigating the distribution of the uncertainties of the fitted modulation amplitudes in each pseudo-experiment ensemble. Based on this analysis, we expect our uncertainty to fall within the interval (0.0037, 0.0041)~dru at a 1$\sigma$ confidence level; we do not observe any dependence of our projected uncertainty on the amplitude of the modulation. Given that the maximum bias observed in this investigation is an order of magnitude smaller than the projected uncertainty, we do not adjust for this bias in our analysis.

In addition to assessing the bias induced by our full model of the event rate over time, we also investigate the bias induced by the simplified background model used in our previous modulation search. This simplified model consisted of a constant and exponentially decaying component in each detector, in addition to the annual modulation signal. The single exponentially decaying component in each detector served to model short-lived cosmogenic isotopes, similar to the exponentially decaying components in the background model described by Eq.~\ref{eq:fit}; however, the specific cosmogenic components present in the detectors were not known at the time of our previous analysis, precluding the use of the more accurate background model presented here. To quantify the bias in the modulation amplitude measured using this simplified background model on the \totalexposure\ dataset studied here, we fit this simplified model to the aforementioned pseudo-experiments. Across all pseudo-experiment ensembles the magnitude of the mean bias is greater than 0.0085~dru, as shown in Fig.~\ref{fig:pseudo}. This large bias, with a magnitude roughly as large as the DAMA-observed modulation amplitude, illustrates the importance of developing an accurate time-dependent background model in annual modulation searches.

\begin{figure}[t]
	\centering
	\includegraphics[width=\columnwidth]{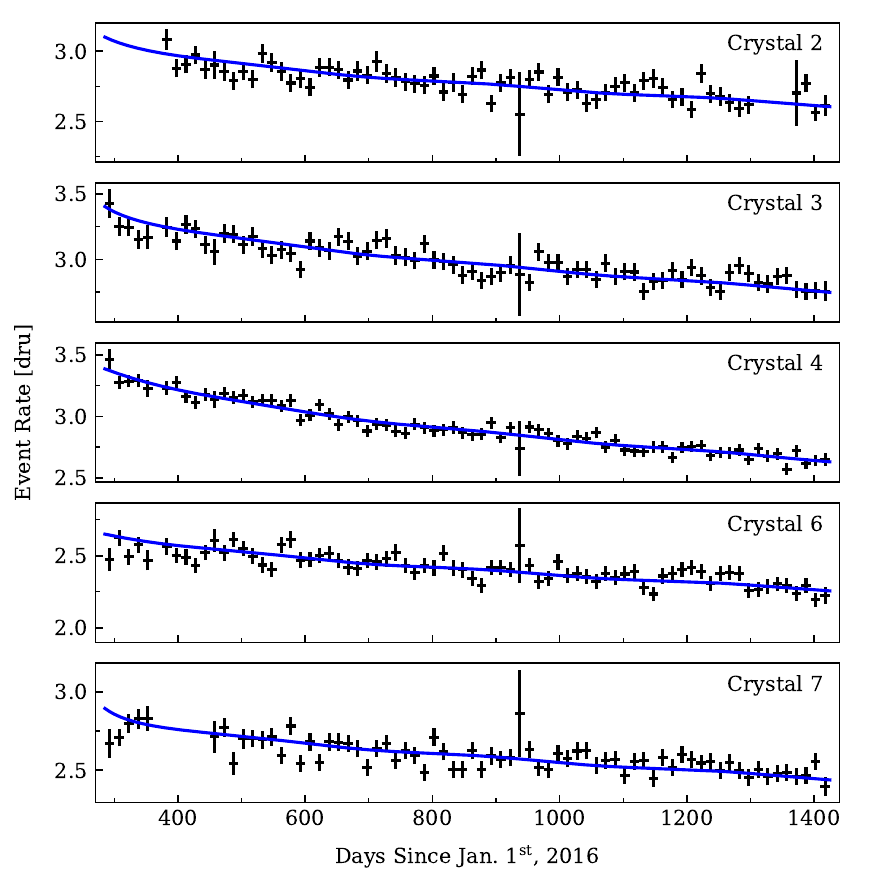}
	\caption{Event rate over time of each detector in the 1--6~keV energy region from \thisbegindate\ to \thisenddate\ binned in 15-day intervals with the phase-fixed best-fit model overlaid.}
    \label{fig:fitted-data}
\end{figure}

\subsection{Results and Discussion}
\label{sec:results}

\begin{figure}[!ht]
	\centering
	\includegraphics[width=\columnwidth]{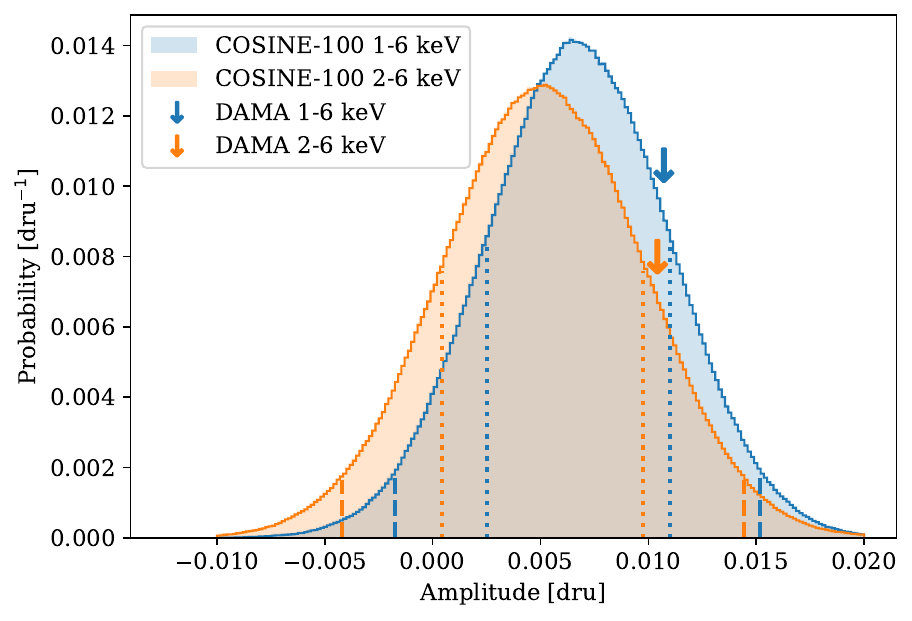}
	\caption{Marginalized posterior distribution of the observed modulation amplitude in the 1--6 and 2--6~keV energy ranges with the modulation phase fixed at June 2\textsuperscript{nd}, 152.5~days from the start of the calendar year. The dotted and dashed lines denote the 68.3\% and 95.5\% highest-density intervals, respectively. The arrows represent the best-fit amplitudes from the results of the DAMA experiments~\cite{DAMAPhase2}.}
    \label{fig:posterior}
\end{figure}

With the phase of the modulation signal fixed at 152.5 days, we find a best-fit modulation amplitude in the 1--6~keV energy region of $0.0067\pm0.0042$~dru. The observed event rate over time overlaid with the phase-fixed best-fit model for this energy region is shown in Fig.~\ref{fig:fitted-data}, and the marginalized posterior distribution of the modulation amplitude in the 1--6~keV region is shown in Fig.~\ref{fig:posterior}. We also perform a fit to our data in the 2--6~keV region and find a best-fit amplitude of $0.0051\pm0.0047$~dru. The measured modulation amplitudes in both energy regions are consistent with both the DAMA-observed modulation and the case of no observed modulation. Table~\ref{tab:results} provides a full summary of the results from this phase-fixed modulation search.

\begin{table*}[!ht]
	\caption{Comparison of modulation search results from \cosine, \mbox{ANAIS-112}, and the DAMA experiments in the 1--6 and 2--6~keV energy intervals for the fit in which the phase and period of the modulation are fixed at 152.5~days and 365.25~days, respectively.}
	\label{tab:results}
    \begin{ruledtabular}
	    \begin{tabular}{lcr}
		    Configuration& Amplitude (dru) & Phase (days)\\
		    \hline
		    COSINE-100 1--6~keV (This result) & 0.0067$\pm$0.0042 & 152.5 (fixed)\\
		    COSINE-100 2--6~keV (This result) & 0.0051$\pm$0.0047 & 152.5 (fixed)\\
		    COSINE-100 2--6~keV (2019 result~\cite{COSINE_modulation}) & 0.0083$\pm$0.0068  & 152.5 (fixed)\\
		    ANAIS 1--6~keV (2021 result~\cite{ANAIS_threeyears})  & -0.0034$\pm$0.0042 & 152.5 (fixed)\\
		    ANAIS 2--6~keV (2021 result~\cite{ANAIS_threeyears})  & 0.0003$\pm$0.0037  & 152.5 (fixed)\\
		    DAMA/LIBRA 1--6~keV (phase2~\cite{DAMAPhase2})             & 0.0105$\pm$0.0011  & 152.5 (fixed)\\
		    DAMA/NaI+LIBRA 2--6~keV~\cite{DAMAPhase2}            & 0.0102$\pm$0.0008  & 152.5 (fixed)\\
	    \end{tabular}
	   \end{ruledtabular}
\end{table*}

\begin{figure}[!ht]
    \centering
    \includegraphics[width=\columnwidth]{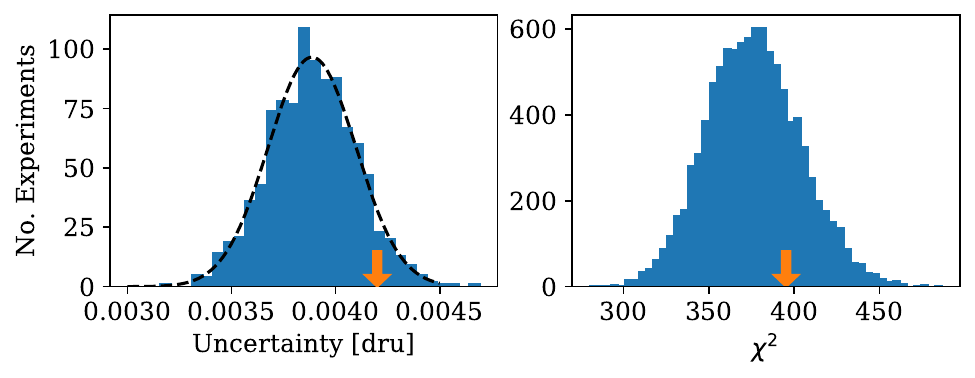}
    \caption{Predicted modulation amplitude uncertainty (left) and chi-square distribution of the best-fit model (right) from pseudo-experiment ensembles for the 1--6 keV phase-fixed modulation search compared with the result from data. Results from the pseudo-experiments are shown as blue histograms, whereas the measured results from data are marked by orange arrows. Here, the uncertainty distribution from the pseudo-experiment ensemble with a 0.0~dru input modulation is shown as a representative example. The black-dashed histogram is the Gaussian fit to the uncertainty distribution.}
    \label{fig:pseudo-sens}
\end{figure}

\begin{figure*}[ht]
	\centering
	\includegraphics[width=\columnwidth]{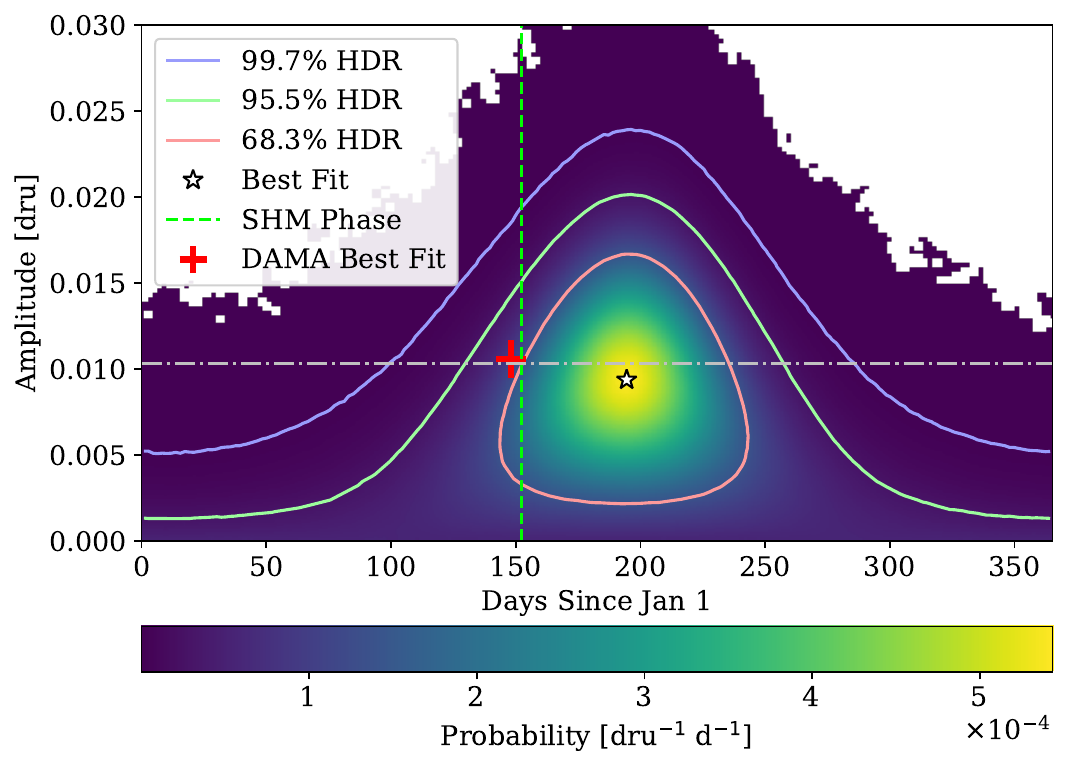}
	\includegraphics[width=\columnwidth]{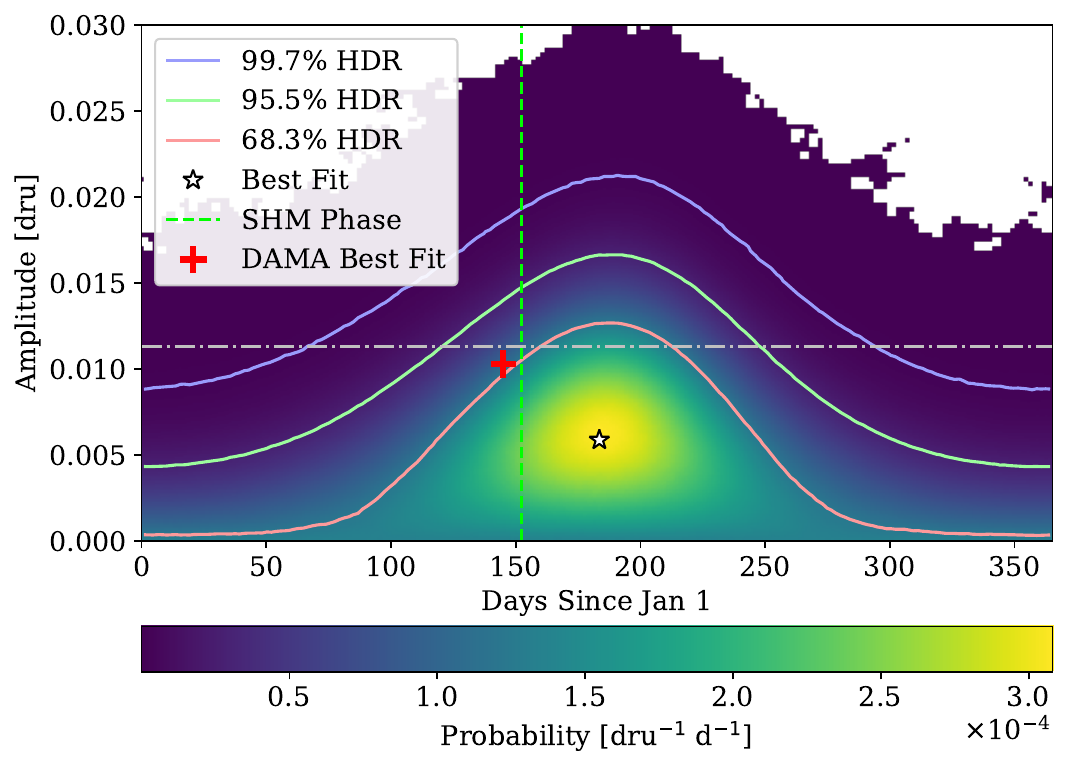}
	\caption{The \cosine\ best-fit point and 68.3\%, 95.5\%, and 99.7\% highest-density credible regions (HDRs) of the modulation amplitude-phase posterior distributions in the 1--6~keV (left) and 2--6~keV (right) energy intervals for the phase-floated modulation search. The best-fit amplitude and phase reported by DAMA~\cite{DAMAPhase2} is included for comparison; the left panel presents the best-fit value from DAMA/LIBRA-phase2 and the right panel presents the combined results from DAMA/NaI and DAMA/LIBRA-phase1+phase2. The silver line denotes the median of the distribution of the maximum modulation amplitude in the 68.3\% HDR from the ensemble of no-modulation pseudo-experiments. The time of maximum dark matter flux expected from the standard halo model (SHM), June 2, or 152.5~days from the start of the calendar year, is also included~\cite{Colloquium}.}
    \label{fig:2Dfit}
\end{figure*}

In judging compatibility of a best-fit model with the data, we do not assume Wilks' theorem~\cite{wilks} to compute the p-value but instead numerically compute the chi-square distribution from ensembles of 10000 pseudo-experiments generated from the best-fit model. As an example, the computed chi-square distribution from the 1--6~keV modulation search and the measured chi-square value of the best-fit model are shown in the right panel of Fig.~\ref{fig:pseudo-sens}. In this case, we find that the best-fit model agrees with the data at a p-value of $p=0.239$. In the 2--6 keV energy region, we find a p-value of $p=0.485$. The data is in good agreement with our best-fit models, as evidenced by the calculated p-values. In addition, the level of precision obtained in these measurements is in agreement with predictions of our uncertainty from the pseudo-experiment analysis described in Sec.~\ref{sec:pseudo}, as can be seen in the left panel of Fig.~\ref{fig:pseudo-sens}.

We also compare the activities of the background components measured in this study with those measured in our search for spin-independent WIMP-nucleus scattering using 1.7~years of data~\cite{cosine_wimp_set2}. We find that the measured activities of both the constant-in-time and shorter-lived background components are in good agreement between the two analyses; however, the measured values of the \isotope{H}{3} and \isotope{Pb}{210} activities are in moderate tension. An initial inspection of this effect finds that the \isotope{Pb}{210} present on the surfaces of the NaI(Tl) crystals~\cite{COSINE_background2} exhibits a decay time that is longer than expected of \isotope{Pb}{210}, while the \isotope{Pb}{210} contained within the bulk of the crystal does not exhibit this effect. An initial assessment of the impact of this effect on the measured modulation amplitude is performed by fitting the measured event rate over time using the model described by Eq.\ref{eq:fit} with an additional longer-lived \isotope{Pb}{210} component added to each NaI(Tl) detector. This updated model returns identical best-fit values of the modulation amplitude to those not including this longer-lived \isotope{Pb}{210} component, while also improving the agreement between the \isotope{H}{3} and \isotope{Pb}{210} background components measured in Ref.~\cite{cosine_wimp_set2} and this analysis. Further investigations are currently underway to confirm the longer-lived \isotope{Pb}{210} component as the cause of the discrepancy between the two analyses, and to confirm the origin of the extended decay time.

The two-dimensional marginalized posterior distributions obtained from the phase-floated modulation search are shown in Fig.~\ref{fig:2Dfit}, which highlights the best-fit points of the model along with the 68.3\%, 95.5\%, and 99.7\% HDRs of the distributions. The best-fit values obtained by the DAMA experiments are also displayed for comparison. It should be noted that the posterior distributions are biased towards positive modulation amplitudes due to the non-linearity of the phase in the fitted model, as described in Ref.~\cite{ANAIS_threeyears}. Nonetheless, we are able to evaluate the level of agreement of the best-fit point for each energy range with the no-modulation case. This is done by comparing the best-fit point of the measured posterior distribution with the posterior distributions of the ensemble of pseudo-experiments with no injected modulation signal. Specifically, we compute the median of the distribution of the maximum amplitude within the  68.3\% HDR of the posterior of each pseudo-experiment, with the computed values shown in Fig.~\ref{fig:2Dfit}. Having taken the effect of this bias into account, we find the results from the phase-floated modulation search are in agreement with both the DAMA-preferred modulation signal and the case of no annual modulation, as in the fixed-phase search.

\begin{figure}[ht]
	\centering
	\includegraphics[width=\columnwidth]{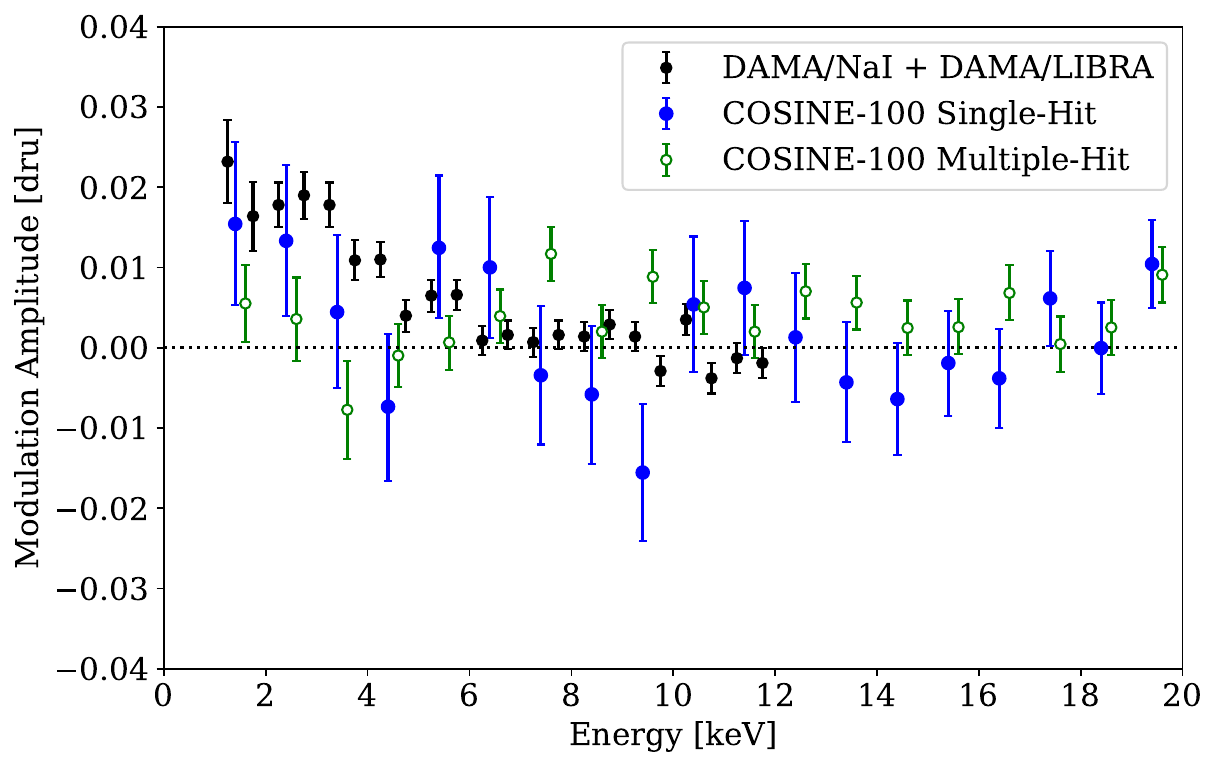}
	\caption{Measured modulation amplitude as a function of energy in 1~keV bins for the \cosine\ single-hit (blue closed circles) and multiple-hit (green open circles) datasets. The combined results from DAMA/NaI and DAMA/LIBRA-phase1 and phase2~\cite{DAMAPhase2} are also shown for reference. The period and phase of the modulation component are fixed at 365.25~days and 152.5~days, respectively. Vertical error bars are the 68.3\% highest-density credible intervals of the modulation amplitude posteriors in each energy bin.}
    \label{fig:amp_v_energy}
\end{figure}

Lastly, we present the best-fit modulation amplitude as a function of energy for both single-hit and multi-hit events in Fig.~\ref{fig:amp_v_energy}. In the two lowest-energy bins, we find a slight preference for the DAMA-observed modulation amplitude, though the no-modulation case also falls within the 95\% HDI of the posterior of both energy bins. This result is consistent with our observations in both the phase-fixed and phase-floated modulation searches in the 1--6 and 2--6~keV energy regions. As described in Sec.~\ref{sec:fit}, no dark matter-induced annual modulation signal is expected in the 6--20~keV single-hit and 1--6~keV multi-hit sideband regions, allowing them to serve as crosschecks of our modulation search procedure. We find that the result in the 6--20~keV single-hit sideband is consistent with no modulation at a level of $\chi^2/$DOF=$12.8/14$. Additionally, the multi-hit 1--6~keV result is consistent with no modulation at $\chi^2/$DOF$=3.53/5$.

\section{Conclusions}
\label{conclusions}

In conclusion, we have performed a search for a dark matter-induced annual modulation signal in NaI(Tl) detectors with \totaluptime\ of data obtained between \thisbegindate\ and \thisenddate\ with \cosine. We have improved upon our previous modulation search~\cite{COSINE_modulation} by implementing a more powerful event selection procedure that allowed us to lower our energy threshold to 1~keV and by implementing a fully featured, time-dependent background model based on dedicated background studies of short-lived components of cosmogenic origin in \cosine. With the phase and period of the modulation signal fixed, we observe a best-fit modulation amplitude of $0.0067\pm0.0042$ ($0.0051\pm0.0047$)~dru in the 1--6 (2--6)~keV signal region, consistent with both the modulation amplitude reported by DAMA and the no-modulation case. In addition, when allowing both the phase and the amplitude of the modulation signal to float as free parameters we measure a best-fit value that is consistent with both the DAMA-preferred value and the case of no modulation. The validity of our modulation search procedure was confirmed via pseudo-experiment studies and analyses of sideband data samples. Although \cosine\ is unable to distinguish between the DAMA-observed modulation and no modulation signal after three years of operation, we plan to continue operation of the \cosine\ detector until at least late 2022, when commissioning for the next phase of the experiment, \mbox{COSINE-200}, is scheduled to commence. Thus, the final exposure of \cosine\ will increase compared with this analysis by more than a factor of two, significantly improving our sensitivity to DAMA's observed modulation signal.

\begin{acknowledgments}
We thank the Korea Hydro and Nuclear Power (KHNP) Company for providing underground laboratory space at Yangyang.
This work is supported by:  the Institute for Basic Science (IBS) under project code IBS-R016-A1 and NRF-2021R1A2C3010989, Republic of Korea;
NSF Grants No. PHY-1913742, DGE-1122492, WIPAC, the Wisconsin Alumni Research Foundation, United States; 
STFC Grant ST/N000277/1 and ST/K001337/1, United Kingdom;
Grant No. 2017/02952-0 FAPESP, CAPES Finance Code 001, CNPq 131152/2020-3, Brazil.

\end{acknowledgments}

\bibliography{ref}

\end{document}